# Few-shot Transfer Learning for Holographic Image Reconstruction using a Recurrent Neural Network

Luzhe Huang[1,2,3], Xilin Yang[1,2,3], Tairan Liu[1,2,3], Aydogan Ozcan[1,2,3,4,*]

1 Electrical and Computer Engineering Department, University of California, Los Angeles, CA 90095, USA
2 Bioengineering Department, University of California, Los Angeles, CA 90095, USA
3 California Nano Systems Institute (CNSI), University of California, Los Angeles, CA 90095, USA
4 David Geffen School of Medicine, University of California, Los Angeles, CA 90095, USA
* Corresponding author: ozcan@ucla.edu

**Abstract**
Deep learning-based methods in computational microscopy have been shown to be powerful but in general face some challenges due to limited generalization to new types of samples and requirements for large and diverse training data. Here, we demonstrate a few-shot transfer learning method that helps a holographic image reconstruction deep neural network rapidly generalize to new types of samples using small datasets. We pre-trained a convolutional recurrent neural network on a large dataset with diverse types of samples, which serves as the backbone model. By fixing the recurrent blocks and transferring the rest of the convolutional blocks of the pre-trained model, we reduced the number of trainable parameters by ~90% compared with standard transfer learning, while achieving equivalent generalization. We validated the effectiveness of this approach by successfully generalizing to new types of samples using small holographic datasets for training, and achieved (i) ~2.5-fold convergence speed acceleration, (ii) ~20% computation time reduction per epoch, and (iii) improved reconstruction performance over baseline network models trained from scratch. This few-shot transfer learning approach can potentially be applied in other microscopic imaging methods, helping to generalize to new types of samples without the need for extensive training time and data.

**Introduction**

The past decade has witnessed rapid advancements stemming from the combination of deep learning and computational imaging fields. For example, deep learning-enabled computational microscopy has been demonstrated on a wide spectrum of imaging tasks, including microscopic image super-resolution[1–3], cross-modality imaging[4–7], volumetric imaging[8–10], and among others[11–15]. As another example, holographic imaging and phase retrieval algorithms have also significantly benefitted from deep learning-based methods to reconstruct complex optical fields from intensity-only measurements, providing improved reconstruction accuracy, speed, and extended depth-of-field[16–20].

However, applying these deep learning-based image reconstruction methods also has some challenges, including e.g., generalization of trained models to new datasets from new types of samples. Due to the limited scale and diversity of available training data, and potential distribution shifts in acquired images, resulting from e.g., varying sample preparation and imaging protocols, a trained neural network model can face inference errors, failing to generalize to new, unknown types of samples. Transfer learning [26] and domain adaptation[27] are two popular few-shot learning methods to help generalize network models on limited labeled data. The first technique fine-tunes a portion or all of the parameters of a pre-trained model on a small dataset with the new distribution, and the second one aims to generalize models on a known target domain during the training without using labels of the target domain. For this, the pre-trained model demands both a high-capacity network that can adapt to different domains well, and a large, diverse training set that represents the distribution variations.

Here, we demonstrate a few-shot transfer learning method for holographic image reconstruction using a recurrent neural network (RNN) that achieves successful generalization to new sample types, never seen during the training. To demonstrate this approach, we pre-trained a convolutional RNN on a large holographic dataset composed of various types of samples (blood smears, Pap smears and lung tissue sections), which served as our backbone model. We show that this backbone model can be rapidly transferred to small training sets of new sample types (e.g., prostate and salivary gland tissue sections – never seen/used before), converging ~2.5x faster compared with baseline models trained from scratch, also saving ~20% training time per epoch using ~90% less number of trainable parameters by fixing the RNN blocks (backbone) in the model. The main contributions of this work include: (1) building a pre-trained holographic image reconstruction model that is suitable for fast few-shot learning on new types of samples, and (2) demonstrating a rapid transfer learning scheme that reduces the training time and the number of trainable parameters by fixing the RNN blocks in the model. We successfully transferred the backbone RNN model to small scale prostate and salivary gland datasets that were never used during the training phase and achieved microscopic reconstruction of inline holograms, correctly revealing the phase and amplitude distributions of these new types of tissue samples with minimal amount of training data and time.

**Results and Discussions**

For holographic image reconstruction we used a convolutional RNN architecture (named RH-M [25]), which was demonstrated to be effective for multi-height phase retrieval and holographic image reconstruction (see Fig. 1a). $M$ intensity-only holograms recorded by a lensfree inline holographic microscope[31] are first back-propagated with zero phase (i.e., without any phase retrieval) by an axial distance of $\bar{z}_2 = 500 \ \mu m$, and then the resulting image sequence is fed into the network as its input (see the Methods section). In the RH-M network, the features in the propagated holograms are extracted by a series of convolutional blocks at different scales, and then aggregated by the RNN blocks (marked by the dash-lined box in Fig. 1a). The ground truth/target sample fields (including sample phase and amplitude) were created by a multi-height phase retrieval (MH-PR) algorithm using 8 separate holograms captured at different sample-to-sensor distances [29]. Here, we pre-trained an RH-M model using a large holographic image dataset including three types of samples: blood smears, pap smears, and lung tissue sections. On average each sample type has $N \cong 700$ non-overlapping, unique fields-of-view (FOVs) and $M = 5$ input holograms were used for reconstruction. However, standard RH-M networks suffer from limited generalization and fail on the reconstruction of entirely new types of samples that were never seen by the network before; see for example the blind testing results of prostate tissue sections in Fig. 1a.

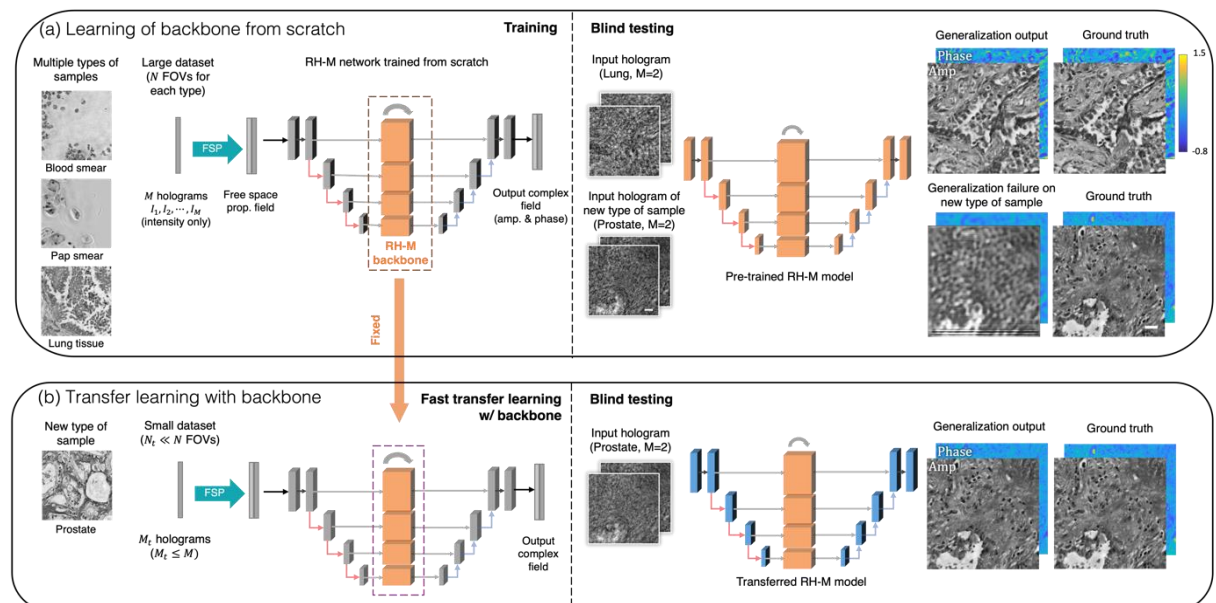

**Figure 1**. The pre-trained RNN backbone model and few-shot transfer learning for holographic image reconstruction. (a) RH-M backbone network pre-trained from scratch on a large dataset (composed of $N$ FOVs for each sample type) with 3 types of samples (blood smears, Pap smears and lung tissue sections). The network is later directly tested on a new type of sample (prostate tissue section) but fails in its reconstruction since this type of sample was never seen by the network before. (b) Transfer learning of an RH-M network from the pre-trained model with fixed RNN backbone. A small dataset of the new sample type with $N_t$ image FOVs is used for transfer learning, where $N_t \ll N$. After a fast transfer learning process, the RH-M can generalize on testing slides of the new sample type very well. Scale bar: 50 μm

Inspired by the fact that the axial differences in hologram intensity patterns reflect/encode the sample's phase information[32], we fixed the RNN blocks (backbone) of the pre-trained RH-M model, which reflect the differences between input holograms and merge their features, and then transferred the rest of the model to new type of samples as shown in Fig. 1b. After a rapid transfer learning process using a small dataset of the new sample type, the resulting new model successfully adapts to the new data distribution of prostate tissue sections and generalizes very well to successfully reconstruct both the phase and amplitude information of the sample (see Figs. 1b and 2).

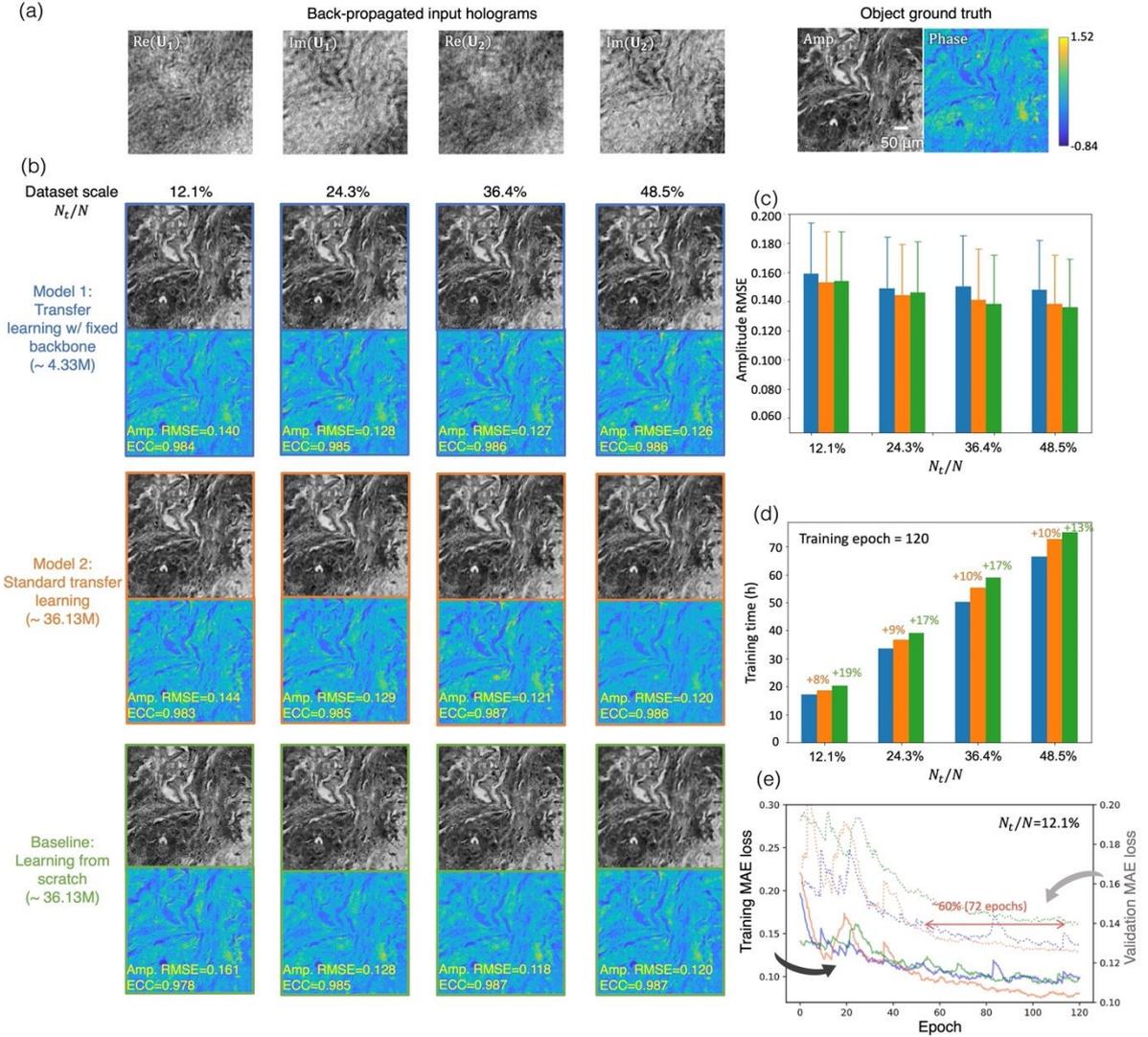

**Figure 2** Transfer learning results of RH-M backbone on prostate tissue sections. Models 1 and 2 were transferred from the pre-trained RH-M model with and without fixed RNN backbone, respectively. Baseline models were trained from scratch. (a) Back-propagated input holograms $U_i$ and ground truth complex field obtained by MH-PR using 8 input holograms. (b) Reconstruction results of models 1, 2 and the baseline model transferred/trained on small training datasets with 4 different scales (i.e., different $\frac{N_t}{N}$ ratios). The numbers in parentheses are the number of trainable parameters for each model. (c) Amplitude root mean square error (RMSE) of the reconstructed complex fields by RH-M models with respect to the ground truth fields, averaged over 49 different FOVs. (d) Training time of RH-M models with increasing training dataset ratio ($\frac{N_t}{N}$). (e) Training and validation loss curves of transferred models and the baseline model on a small dataset with $\frac{N_t}{N} = 12.1\%$.

Major advantages of our approach resulting from the generalization of a pre-trained model in transfer learning include faster convergence speed and better adaptability to small training datasets. To better quantify these advantages, we evaluated the transfer learning performance of the backbone model using a series of datasets with different $\frac{N_t}{N}$ ratios. As illustrated in Fig. 2, the pre-trained model was transferred onto prostate datasets using different amounts of unique sample FOVs, i.e., $N_t$. We created models using the reported transfer learning scheme with a fixed backbone (model 1) and standard transfer learning (without fixed backbone, model 2) on 4 training datasets of various $\frac{N_t}{N}$ ratios; each input sequence contained $M_t = 2$ back-propagated holograms during the transfer. We additionally trained the same network from scratch (baseline model) on the same 4 datasets with different $\frac{N_t}{N}$ for comparison. After convergence (about 120 epochs), we tested them on an additional testing dataset of a prostate tissue excluded from all training sets. As indicated in Fig. 2b, models 1 and 2 showed equivalent reconstruction performance compared with the baseline model on all 4 datasets with different $\frac{N_t}{N}$. Increasing the ratio of the training dataset benefited the reconstruction quality as further indicated by the decreasing amplitude RMSE and the increasing enhanced correlation coefficient (ECC) values (see the Methods section). Figure 2c further confirmed the same conclusion by calculating the amplitude RMSE of the model outputs on the testing set of 49 unique FOVs.

Furthermore, as noted in Fig. 2b, the reported approach (model 1) used only ~4.33 million trainable parameters, compared to >36 million for the other models, saving ~90% of the trainable parameters. Another advantage of the reported approach is the reduced time cost of transfer learning. As reported in Fig. 2d, the training time of our approach (model 1, blue bars) was reduced by up to 19% on the same GPU machine (see Methods) compared to standard transfer learning (model 2, orange bars) and the baseline model (green bars). Figure 2e further compared the training and validation mean absolute error (MAE) values for the transferred models and the baseline model on a small prostate tissue section dataset (with $\frac{N_t}{N} = 12.1\%$). The transferred models (model 1 and 2, blue and orange curves respectively) both converged ~2.5x faster than the baseline model (green curves), saving about 60% of the training epochs to reach the same performance in terms of the validation MAE loss.

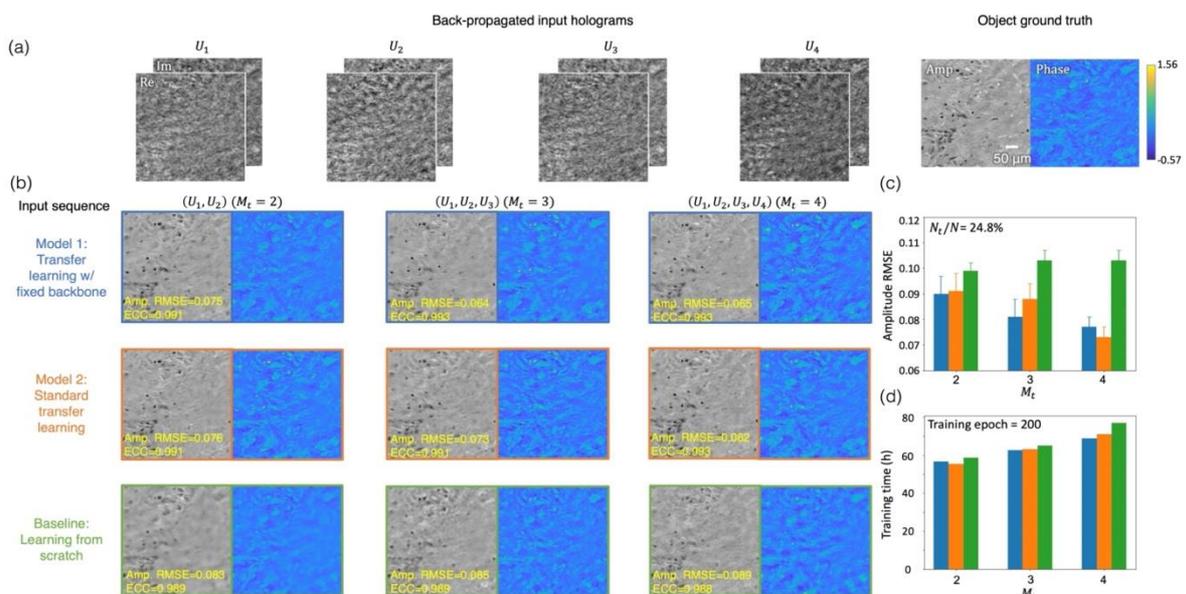

**Figure 3** Transfer learning results of RH-M backbone on salivary gland tissue sections. Models 1 and 2 were transferred from the pre-trained RH-M model with and without fixed RNN backbone, respectively. Baseline model was trained from scratch. (a) Back-propagated input holograms $U_i$ and the ground truth field obtained by MH-PR from 8 input holograms. (b) Reconstruction results of models 1, 2 and the baseline model transferred/trained on small training datasets with different numbers of input holograms

($M_t = 2,3,4$). (c) Amplitude RMSE of the reconstructed fields by RH-M models with respect to the ground truth fields, averaged over 40 unique testing input – ground truth pairs; (d) Training time of the transferred models and the baseline model on datasets with different $M_t$.

Next, we evaluated the generalization of the backbone model to different input sequence lengths $M_t$, using an additional, new sample type. The training set was captured on a few salivary gland tissue sections ($\frac{N_t}{N} = 24.8\%$), and the backbone model was transferred to datasets with $M_t = 2,3,4$ holograms respectively, with and without a fixed backbone. In addition, baseline models with $M_t = 2,3,4$ were also trained from scratch (using the same amount of data) for comparison purposes. Then, the models were blindly tested using another testing set (Fig. 3a). As shown in Fig. 3b, transferred models (models 1, 2) successfully reconstructed the sample complex field with high fidelity, reflected by the low amplitude RMSE and high ECC. Furthermore, by adding extra input holograms, i.e., increasing $M_t$, the reconstruction accuracy can be further enhanced. In contrast, the output images of the baseline models (labeled green) are severely contaminated by artifacts (due to limited amount of training data available for the new sample type), scoring worse amplitude RMSE and ECC values than those achieved by our transferred models. Figure 3c further illustrates the amplitude RMSE values of these models' outputs on an external testing set with 40 input – ground truth pairs. For all $M_t$ values, the transferred models (blue and orange bars) significantly outperform the baseline models (green bars). Both Figs. 3b and c indicate that the reconstruction accuracy of the transferred models can be further improved by increasing $M_t$, confirming the effectiveness of the pre-trained RNN backbone in multi-height holographic image reconstruction. On the other hand, baseline models trained from scratch failed to learn using small training sets and cannot efficiently utilize additional holograms, resulting in a flat trend of the amplitude RMSE values (green bars in Fig. 3c). Figure 3d further compares the training time of the transferred models and the baseline model with respect to $M_t$, demonstrating the fast generalization of the backbone model.

**Conclusions**
In this work, in order to improve deep neural networks' generalization to successfully image new sample types in computational holographic microscopy, we presented a transfer learning-based few-shot learning method to generalize models when only small datasets of new sample types are available; we demonstrated the success of this new approach using a convolutional RNN to perform multi-height holographic image reconstruction. We established an RNN backbone model and a transfer learning scheme to reduce both the computation time and the number of trainable parameters compared to standard transfer learning approaches. The generalization of the backbone model was validated on prostate and salivary gland tissue datasets that were never seen by the network before. Compared with baseline models trained from scratch, the RNN models transferred from the backbone gained faster convergence speed and improved the image reconstruction quality. The reported transfer learning framework substantially enhances the generalization of deep neural network-based holographic imaging for new types of samples.

**Methods**
**Imaging system and samples**
Experiments in this work were implemented using a lens-free in-line holographic microscope. A broadband light source (WhiteLase Micro, NKT Photonics) was used for illumination, filtered by an acousto-optic tunable filter ($530 nm$). A complementary metal-oxide-semiconductor (CMOS) image sensor (IMX 081, Sony) captures the raw holograms. A 3D positioning stage (MAX 606, Thorlabs, Inc.,) was used to move the CMOS sensor to perform precise lateral and axial shifts. The samples of interest were directly placed between the light source and the CMOS sensor. The typical sample-to-source distance ($z_1$) and sample-to-sensor distance ($z_2$) used for imaging ranged from $\sim 5 - 10$ cm and $\sim 400 - 600$ μm respectively. All hardware was controlled by a customized LabVIEW program during the imaging process.

All human samples involved in this work were deidentified and prepared from existing specimens, without a link or identifier to the patient. Human prostate and lung tissue slides were prepared by and acquired from the UCLA Translational Pathology Core Laboratory (TPCL). Pap smear slides were provided by the UCLA Department of Pathology. Blood smear slides were provided by the UCLA Department of Internal Medicine.

The raw in-line holograms were algorithmically super-resolved. To implement pixel super-resolution, 6-by-6 inline holograms were captured for each FOV with subpixel shifts using a 3D positioning stage (MAX606, Thorlabs, Inc.). Relative lateral shifts were first estimated by an image correlation-based algorithm, and then 36 holograms were shifted and added to obtain a super-resolved hologram[29]. The resulting super-resolved holograms were used for both multi-height phase retrieval and neural network-based hologram reconstructions.

**Multi-height phase recovery and free space propagation**
8 in-line holograms at different sample-to-sensor distances were captured to perform MH-PR for each sample FOV[29]. An autofocusing algorithm is first applied to estimate the sample-to-sensor distance for each hologram based on the edge sparsity criterion[33]. Then the first hologram with zero phase padding is propagated to the remaining hologram planes, using the angular spectrum-based wave propagation[34] and the estimated sample-to-sensor distances. At the designated hologram position, the resulting field is updated using the measured hologram at the same position, where the amplitude of the resulting field is averaged with the measured hologram amplitude and the phase is kept/retained. One iteration is completed when all the measured holograms have been used in a sequence, and this iterative algorithm typically converges after 10-30 iterations. Finally, the resulting field is back-propagated to the sample plane to retrieve its phase and amplitude images.

**Image quality evaluation metrics**
RMSE, ECC, MAE and multiscale structural similarity index (SSIM) were used to evaluate the similarity between two images during the training and testing of the presented neural networks. Denoting the two images as $x$ and $y$ with dimensions of $H \times W$, these metrics are defined as follows:

$$RMSE(x,y) = \sqrt{\frac{1}{HW}\sum_{i=1}^{H}\sum_{j=1}^{W}(x(i,j)-y(i,j))^2}$$

$$ECC(x,y) = \frac{Re\{\sum_{i=1}^{H}\sum_{j=1}^{W}x^*(i,j)y(i,j)\}}{\sqrt{\left(\sum_{i=1}^{H}\sum_{j=1}^{W}|x(i,j)|^2\right)\cdot\left(\sum_{i=1}^{H}\sum_{j=1}^{W}|y(i,j)|^2\right)}}$$

$$MAE(x,y) = \sum_{i=1}^{H}\sum_{j=1}^{W}|x(i,j) - y(i,j)|$$

$$SSIM(x,y) = \left(\frac{2\mu_x\mu_y + C_1}{\mu_x^2 + \mu_y^2 + C_1}\right)^{\alpha_M} \prod_{s=1}^{M}\left(\frac{2\sigma_{x,s}\sigma_{y,s} + C_2}{\sigma_{x,s}^2 + \sigma_{y,s}^2 + C_2}\right)^{\beta_s}\left(\frac{\sigma_{xy,s} + C_3}{\sigma_{x,s}\sigma_{y,s} + C_3}\right)^{\gamma_s}$$

where $i, j$ are indices of image pixels. $\mu_{x,s}$ and $\sigma_{x,s}$ are the mean and standard deviation values of $2^{s-1}$ downsampled version of $x$, respectively, and $\sigma_{xy,s}$ is the covariance between $2^{s-1}$ downsampled versions of $x$ and $y$. The other parameters used for the SSIM calculations are empirically determined as $\beta_1 = \gamma_1 = 0.0448, \beta_2 = \gamma_2 = 0.2856, \beta_3 = \gamma_3 = 0.3001, \beta_4 = \gamma_4 = 0.2363, \alpha_5 = \beta_5 = \gamma_5 = 0.1333, M = 5, C_1 = (0.01L)^2, C_2 = 2C_3 = (0.03L)^2$, where $L = 255$ for 8-bit gray scale images[35]. ECC values are calculated on complex images using both amplitude and phase channels.

**Network architecture and implementation details**

Holograms and their corresponding ground truth fields were cropped into non-overlapping $512 \times 512$-pixel image patches, each corresponding to a $\sim 0.2 \times 0.2\ mm^2$ unique sample FOV. Then the image pair sets of each sample type were divided into training, validation, and testing sets, where the testing set was strictly captured on a different patient slide not used in training and validation sets. During the transfer learning, data augmentation techniques (flipping and rotation) were applied to expand the training set by 8 times. The holographic image reconstruction network (RH-M) follows the convolutional recurrent neural network architecture as in Ref. [8,25]. The down- and up-sampling paths of RH-M consist of 4 consecutive convolutional blocks respectively and 4 RNN blocks connecting the corresponding convolutional blocks in the down- and up-sampling paths. The RNN block adapts two convolutional gated recurrent unit (GRU)[36] layers and one $1 \times 1$ convolutional layer.

A generative adversarial network (GAN)[37] framework was employed in this work. The loss function used for both training and transfer learning is a linear combination of (1) pixel-wise MAE loss: $L_{MAE}$, (2) multiscale SSIM loss, $L_{SSIM}$, between the network output $\hat{y}$ and the ground truth image $y$, and (3) the adversarial loss, $L_{adv}$, given by the discriminator ($D$) network. Accordingly, the total loss for the generator (RH-M) can be expressed as:

$$L_G = \alpha L_{MAE} + \beta L_{SSIM} + \gamma L_{adv}$$

where $\alpha, \beta, \gamma$ are empirically set as 3, 1, 0.3 for all models. MAE and SSIM losses are defined as:

$$L_{MAE} = MAE(y,\hat{y}), \qquad L_{SSIM} = 1 - SSIM(y,\hat{y})$$

Squared loss terms were employed for the adversarial loss and the total discriminator loss ($L_D$) [38]:

$$L_{adv} = [D(\hat{y}) - 1]^2$$
$$L_D = \frac{1}{2}D(\hat{y})^2 + \frac{1}{2}[D(y) - 1]^2$$

Adam optimizers [39] with learning rates of $10^{-5}$ and $10^{-6}$ were used for the generator and discriminator networks, respectively, in the backbone training. Decaying learning rates with initial values of $2 \times 10^{-4}$ and $2 \times 10^{-5}$ were applied for transfer learning on new datasets for the generator and the discriminator, respectively. Based on the validation losses, models are early stopped at 120 and 200 epochs for prostate and salivary gland samples respectively. All models were realized in TensorFlow on a computer with an Intel Xeon W-2195 processor and four NVIDIA RTX 2080 Ti GPUs.

**References**

1. Ouyang, W., Aristov, A., Lelek, M., Hao, X. & Zimmer, C. Deep learning massively accelerates super-resolution localization microscopy. *Nat Biotechnol* **36**, 460–468 (2018).
2. Wang, H. *et al.* Deep learning enables cross-modality super-resolution in fluorescence microscopy. *Nat Methods* **16**, 103–110 (2019).
3. Qiao, C. *et al.* Evaluation and development of deep neural networks for image super-resolution in optical microscopy. *Nat Methods* **18**, 194–202 (2021).
4. Rivenson, Y. *et al.* Virtual histological staining of unlabelled tissue-autofluorescence images via deep learning. *Nat Biomed Eng* **3**, 466–477 (2019).
5. Christiansen, E. M. *et al.* In Silico Labeling: Predicting Fluorescent Labels in Unlabeled Images. *Cell* **173**, 792-803.e19 (2018).
6. de Haan, K. *et al.* Deep learning-based transformation of H&E stained tissues into special stains. *Nat Commun* **12**, 4884 (2021).
7. Cheng, S. *et al.* Single-cell cytometry via multiplexed fluorescence prediction by label-free reflectance microscopy. *Sci. Adv.* **7**, eabe0431 (2021).
8. Huang, L., Chen, H., Luo, Y., Rivenson, Y. & Ozcan, A. Recurrent neural network-based volumetric fluorescence microscopy. *Light Sci Appl* **10**, 62 (2021).
9. Nehme, E. *et al.* DeepSTORM3D: dense 3D localization microscopy and PSF design by deep learning. *Nat Methods* **17**, 734–740 (2020).
10. Yang, X. *et al.* Deep-Learning-Based Virtual Refocusing of Images Using an Engineered Point-Spread Function. *ACS Photonics* **8**, 2174–2182 (2021).
11. Nguyen, T., Xue, Y., Li, Y., Tian, L. & Nehmetallah, G. Deep learning approach for Fourier ptychography microscopy. *Opt. Express* **26**, 26470 (2018).
12. Barbastathis, G., Ozcan, A. & Situ, G. On the use of deep learning for computational imaging. *Optica* **6**, 921–943 (2019).
13. de Haan, K., Rivenson, Y., Wu, Y. & Ozcan, A. Deep-Learning-Based Image Reconstruction and Enhancement in Optical Microscopy. *Proc. IEEE* **108**, 30–50 (2020).
14. Pinkard, H., Phillips, Z., Babakhani, A., Fletcher, D. A. & Waller, L. Deep learning for single-shot autofocus microscopy. *Optica* **6**, 794 (2019).
15. Bostan, E., Heckel, R., Chen, M., Kellman, M. & Waller, L. Deep phase decoder: self-calibrating phase microscopy with an untrained deep neural network. *Optica* **7**, 559 (2020).
16. Goy, A., Arthur, K., Li, S. & Barbastathis, G. Low Photon Count Phase Retrieval Using Deep Learning. *Phys. Rev. Lett.* **121**, 243902 (2018).
17. Wang, H., Lyu, M. & Situ, G. eHoloNet: a learning-based end-to-end approach for in-line digital holographic reconstruction. *Opt. Express* **26**, 22603–22614 (2018).
18. Ren, Z., Xu, Z. & Lam, E. Y. End-to-end deep learning framework for digital holographic reconstruction. *Adv. Photon.* **1**, 016004 (2019).
19. Wang, K., Dou, J., Kemao, Q., Di, J. & Zhao, J. Y-Net: a one-to-two deep learning framework for digital holographic reconstruction. *Opt. Lett.* **44**, 4765–4768 (2019).
20. Deng, M., Li, S., Goy, A., Kang, I. & Barbastathis, G. Learning to synthesize: robust phase retrieval at low photon counts. *Light Sci Appl* **9**, 36 (2020).

21. Rivenson, Y., Zhang, Y., Günaydın, H., Teng, D. & Ozcan, A. Phase recovery and holographic image reconstruction using deep learning in neural networks. *Light Sci Appl* **7**, 17141 (2018).
22. Wu, Y. *et al.* Extended depth-of-field in holographic imaging using deep-learning-based autofocusing and phase recovery. *Optica* **5**, 704–710 (2018).
23. Wu, Y. *et al.* Bright-field holography: cross-modality deep learning enables snapshot 3D imaging with bright-field contrast using a single hologram. *Light Sci Appl* **8**, 25 (2019).
24. Rivenson, Y., Wu, Y. & Ozcan, A. Deep learning in holography and coherent imaging. *Light Sci Appl* **8**, 85 (2019).
25. Huang, L. *et al.* Holographic Image Reconstruction with Phase Recovery and Autofocusing Using Recurrent Neural Networks. *ACS Photonics* **8**, 1763–1774 (2021).
26. Pan, S. J. & Yang, Q. A Survey on Transfer Learning. *IEEE Trans. Knowl. Data Eng.* **22**, 1345–1359 (2010).
27. Tzeng, E., Hoffman, J., Saenko, K. & Darrell, T. Adversarial Discriminative Domain Adaptation. *arXiv:1702.05464 [cs]* (2017).
28. Greenbaum, A., Sikora, U. & Ozcan, A. Field-portable wide-field microscopy of dense samples using multi-height pixel super-resolution based lensfree imaging. *Lab Chip* **12**, 1242 (2012).
29. Greenbaum, A. & Ozcan, A. Maskless imaging of dense samples using pixel super-resolution based multi-height lensfree on-chip microscopy. *Opt. Express* **20**, 3129–3143 (2012).
30. Rivenson, Y. *et al.* Sparsity-based multi-height phase recovery in holographic microscopy. *Sci Rep* **6**, 37862 (2016).
31. Greenbaum, A. *et al.* Imaging without lenses: achievements and remaining challenges of wide-field on-chip microscopy. *Nat Methods* **9**, 889–895 (2012).
32. Kou, S. S., Waller, L., Barbastathis, G. & Sheppard, C. J. R. Transport-of-intensity approach to differential interference contrast (TI-DIC) microscopy for quantitative phase imaging. *Opt. Lett.* **35**, 447–449 (2010).
33. Zhang, Y., Wang, H., Wu, Y., Tamamitsu, M. & Ozcan, A. Edge sparsity criterion for robust holographic autofocusing. *Opt. Lett.* **42**, 3824–3827 (2017).
34. Goodman, J. W. *Introduction to Fourier optics*. (Roberts & Co, 2005).
35. Wang, Z., Simoncelli, E. P. & Bovik, A. C. Multiscale structural similarity for image quality assessment. in *The Thrity-Seventh Asilomar Conference on Signals, Systems & Computers, 2003* 1398–1402 (IEEE, 2003). doi:10.1109/ACSSC.2003.1292216.
36. Cho, K. *et al.* Learning Phrase Representations using RNN Encoder-Decoder for Statistical Machine Translation. *arXiv:1406.1078 [cs, stat]* (2014).
37. Goodfellow, I. J. *et al.* Generative Adversarial Networks. *arXiv:1406.2661 [cs, stat]* (2014).
38. Mao, X. *et al.* Least Squares Generative Adversarial Networks. *arXiv:1611.04076 [cs]* (2017).
39. Kingma, D. P. & Ba, J. Adam: A Method for Stochastic Optimization. *arXiv:1412.6980 [cs]* (2017).